\documentclass{article}
\usepackage[T1]{fontenc} 
\usepackage[utf8]{inputenc} 
\usepackage{ismir,amsmath,cite,url}
\usepackage{graphicx}
\usepackage{color}
\DeclareMathOperator*{\argminB}{argmax}
\usepackage{amssymb}
\usepackage{subcaption}
\usepackage{xcolor}
\usepackage{soul}

\usepackage{lineno}

\title{I can listen but cannot read: \\ An evaluation of two-tower multimodal systems \\ for instrument recognition}

\threeauthors
  {Yannis Vasilakis} {Queen Mary University of London \\ {\tt i.vasilakis@qmul.ac.uk}}
  {Rachel Bittner} {Spotify \\ {\tt rachelbittner@spotify.com}}
  {Johan Pauwels} {Queen Mary University of London \\ {\tt j.pauwels@qmul.ac.uk}}

\def\authorname{Y. Vasilakis, R. Bittner, and J. Pauwels}

\begin{document}

\maketitle
\begin{abstract}
Music two-tower multimodal systems integrate audio and text modalities into a joint audio-text space, enabling direct comparison between songs and their corresponding labels. These systems enable new approaches for classification and retrieval, leveraging both modalities. Despite the promising results they have shown for zero-shot classification and retrieval tasks, closer inspection of the embeddings is needed. This paper evaluates the inherent zero-shot properties of joint audio-text spaces for the case-study of instrument recognition. We present an evaluation and analysis of two-tower systems for zero-shot instrument recognition and a detailed analysis of the properties of the pre-joint and joint embedding spaces. Our findings suggest that audio encoders alone demonstrate good quality, while challenges remain within the text encoder or joint space projection. Specifically, two-tower systems exhibit sensitivity towards specific words, favoring generic prompts over musically informed ones. Despite the large size of textual encoders, they do not yet leverage additional textual context or infer instruments accurately from their descriptions. Lastly, a novel approach for quantifying the semantic meaningfulness of the textual space leveraging an instrument ontology is proposed. This method reveals deficiencies in the systems' understanding of instruments and provides evidence of the need for fine-tuning text encoders on musical data.

\end{abstract}

\section{Introduction}\label{sec:introduction}
Multiclass classification has been a heavily researched topic in Music Information Retrieval (MIR) with many concrete applications such as genre, instrument and emotion recognition~\cite{genre_classification_review_ndou_2021, music_emotion_recognition_survey_donghong_2022, music_classification_beyond_supervised_learning_won_2021, a_survey_music_classification_2011}. Despite the success of Deep (DL) systems for such tasks, recurring deficiencies persist among these problems. These are: (1) the limited availability of large-scale annotated datasets curated by experts, (2) the restricted capability of these systems to infer only a set of predefined classes.

\begin{figure}
 \centerline{\includegraphics[width=\columnwidth, height=3.5cm]{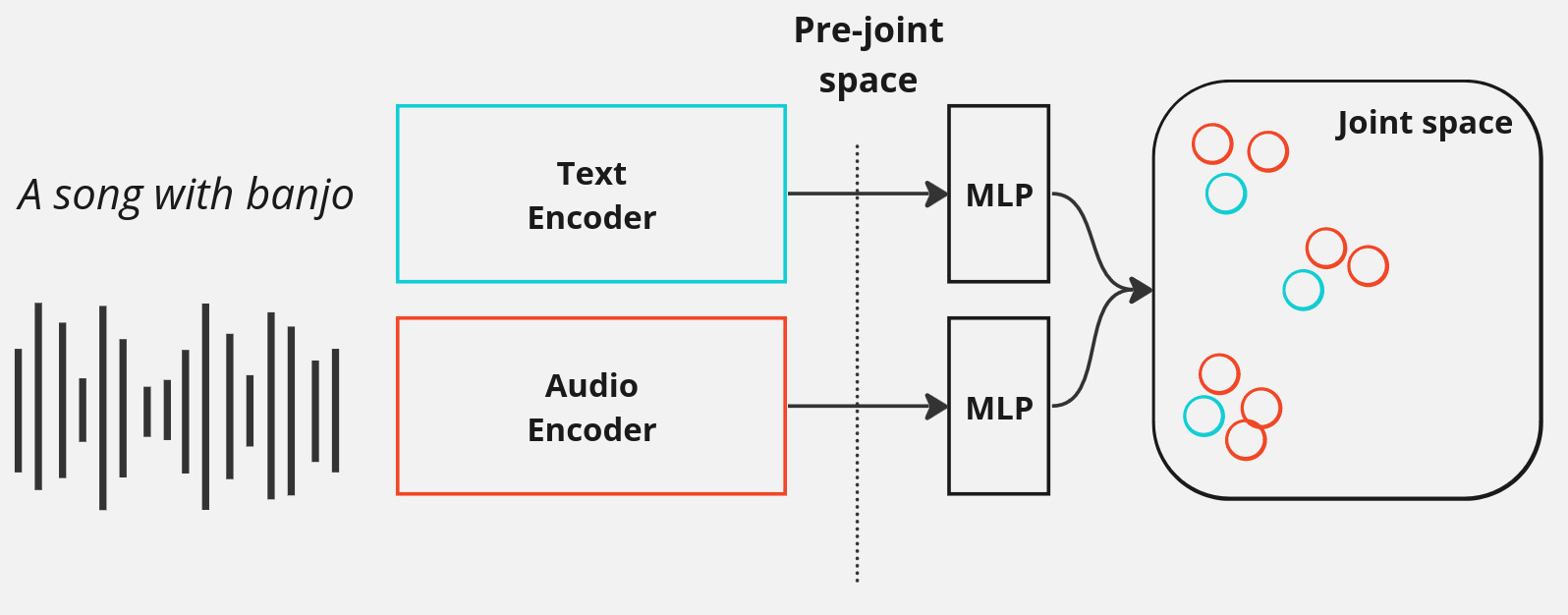}}
 \caption{Figure of a pipeline for two-tower multimodal systems. A separate model for each modality is used and their individual representations are projected to a joint audio-text space through a Multi-Layer Perceptron (MLP). This enables direct comparison between audio and textual data. We refer to embeddings obtained before joint-space projection as pre-joint space embeddings.}
 \label{fig:abstract_two_tower_system}
\end{figure}

Music is an ever-evolving art form and as a result, there is an inherent need to make these systems adaptable to new terms/classes~\cite{classification_as_culture_jennifer_2008, categorical_ambiguity_genre_Venrooij_2018, Hennequin2018AudioBD} and infer task-agnostic representations that can be useful for a plethora of downstream tasks with representation learning \cite{pann_kong_2019, hts_at_a_hierarchical_Ke_2022, towards_learning_universal_audio_rep_wang_2021}. Zero-shot learning (ZSL) is focused on estimating a classifier capable of inferring unseen, new classes without annotated examples~\cite{zero_shot_factored_linear_huang_2020, zero_shot_learning_for_audio_choi_2019, zeroshot_semantic_embeddings_huang_2020, zershot_class_label_embeddings_2019}. ZSL is often achieved in either of two ways: (1) by decomposing each class into attributes and inferring unseen classes through their related attributes (e.g. genres decomposed into presence or absence of instruments~\cite{zero_shot_learning_for_audio_choi_2019}) or (2) using word embeddings from Language Models (LM)~\cite{zero_shot_learning_for_audio_choi_2019, zeroshot_semantic_embeddings_huang_2020, zershot_class_label_embeddings_2019}. The success of contextualized Large Language Models (LLM) has driven the research community predominantly toward the second solution, as it doesn't require experts to define attributes and the mapping between classes and attributes~\cite{Pratt_2023_ICCV, Ge_2023_CVPR, Goyal_2023_CVPR}.

As a result, ZSL for audio classification is primarily focused on connecting the audio and semantic representation spaces. This interconnection can happen in 2 main ways: (1) mapping the audio representations to text space~\cite{zeroshot_semantic_embeddings_huang_2020, zero_shot_learning_for_audio_choi_2019}, or (2) mapping both of the spaces to a new, joint audio-text space~\cite{zershot_class_label_embeddings_2019, muscall_manco_2022, mulan_a_joint_embedding_qinqing_2022, clap_learning_audio_concepts_elizalde_2023}. The systems of the second category are named two-tower multimodal systems, where pre-trained audio models and LMs are used as the audio and text encoders respectively. Representations obtained from each modality are then mapped to a joint audio-text space and systems are jointly optimized such that the audio and text representation are close in the joint space (e.g. the phrase ``A rock song track'' is similar to the recording of a ``rock'' song). We will call such representations as embeddings from here on.

This work aims to better understand the properties of existing two-tower systems. We use instrument classification as a case-study to provide insights into the presence (or absence) of semantic properties in the audio, text or joint spaces in addition to reporting classification metrics. Concretely, we consider 3 systems: MusCALL~\cite{muscall_manco_2022}, a CLAP~\cite{clap_learning_audio_concepts_elizalde_2023, laion_clap_Wu_2023} model trained on speech and music datasets, and a CLAP model trained on music data~\cite{laion_clap_Wu_2023}. We evaluate the performance of these systems on instrument classification using the TinySOL dataset~\cite{TinySOL}. We would like to highlight that multimodal DL models typically excel at simple tasks and datasets like this.

Furthermore, a novel approach for quantifying the semantic meaningfulness of textual encoders for instrument recognition is proposed.

For reproducibility, our experiments are performed on open-source datasets and the code of our experiments is made publicly available\footnote{\url{https://github.com/YannisBilly/i_can_listen_but_cannot_read}}, such that they can be reproduced.
\section{Related work}\label{sec:related_work}

\subsection{Zero-shot transfer}\label{subsec:zeroshot_learning}

ZSL focuses on estimating classifiers for novel, unseen classes without annotated examples. Two-tower systems are not primarily optimized for ZSL but due to the pretrained textual encoder, novel words or phrases can be interpreted during inference. This property is known as zero-shot transfer (ZST)~\cite{lit_zero_shot_transfer_Zhai_2022}.

Side-information can be used in multiple ways that fall into two categories: (1) decomposing classes into shared attributes and (2) using LMs to represent this information as a text embedding. Despite its success, the first solution requires experts to effectively estimate the relevance of attributes and several classes and is a costly activity. As a result and due to the remarkable results obtained through contextualized LLMs, research has focused on the second option.

Generally,  the methodology can be broken down into 3 components: (1) an audio encoder, (2) a textual encoder and (3) a projection to a common space. General purpose audio DL models that have been used as the audio encoder include VGGish~\cite{vggish_simonyan_2014}, PANN~\cite{pann_kong_2019}, HTS-AT~\cite{hts_at_a_hierarchical_Ke_2022} and Audio Spectrogram Transformers~\cite{audio_spectrogram_transformer_2021}. For the textual encoder, distributional LMs like GloVE~\cite{glove_pennington_2014}, Word2Vec~\cite{word2vec_mikolov_2013} and contextualized LMs like BERT~\cite{bert_devlin_2019} have been thoroughly tested. As each modality produces heterogeneous representations, different methods of establishing comparability have been tested. This is predominantly achieved through projecting audio to text/semantic space~\cite{zero_shot_factored_linear_huang_2020, zero_shot_learning_for_audio_choi_2019, zeroshot_semantic_embeddings_huang_2020} or a novel, joint audio-text space \cite{muscall_manco_2022, mulan_a_joint_embedding_qinqing_2022, clap_learning_audio_concepts_elizalde_2023}.

\subsection{Two-tower multimodal systems}\label{subsec:two_tower_multimodal_systems}
Multimodal systems aim to represent data with additional knowledge from multiple modalities. Examples are audio combined with images~\cite{zero_shot_audio_image_dogan_2022}, text~\cite{zero_shot_learning_for_audio_choi_2019, Du2023JointMA} or a combination thereof.

Two-tower multimodal systems focus on combining the textual and audio modalities by projecting them in a joint audio-text space. In that space, words that are relevant to a specific song will produce embeddings that will be close in terms of some similarity metric. An illustration of a two-tower system is presented in Figure \ref{fig:abstract_two_tower_system}. Information flowing through the audio encoder or textual encoder is referred to as the audio and textual branch respectively. We are also interested in the embeddings obtained through the encoders before projecting them into the joint audio-text space. We will call these the pre-joint spaces from now on.

The text used during training is usually a description of a song and will be referred to as a caption. The text used during inference will be referred to as a prompt. 

MusCALL~\cite{muscall_manco_2022} combined a ResNET-50 for audio~\cite{resnet_50_2016} with a Transformer for text encoding~\cite{Vaswani2017AttentionIA}, optimized jointly over InfoNCE contrastive loss~\cite{oord2018representation}. Additionally, a weighing mechanism based on caption-caption similarity was incorporated between negative audio-caption pairs. This is based on the premise that similar captions will be given to similar audio. The audio used is private but the training code is publicly available and used for this work.

MuLan~\cite{mulan_a_joint_embedding_qinqing_2022} experimented with ResNET-50 as well as Audio Spectrogram Transformers~\cite{audio_spectrogram_transformer_2021} for audio encoding and a pretrained BERT model for the text branch. Both were jointly optimized over the Contrastive Multiview Coding loss~\cite{Tian2019ContrastiveMC}, which is a cross-modal extension of InfoNCE. Neither the data nor the code is available.

LAION-CLAP tested 6 different combinations of audio and text embedding models, the best one of which was HTS-AT with RoBERTa~\cite{Liu2019RoBERTaAR}. The latter is the one that will be used in this paper. The LAION-Audio-630k dataset was formed by combining AudioCaps, CLotho and Audioset.

Generally, research for two-tower systems is limited to testing different combinations of audio and text encoders, optimized jointly over a form of contrastive loss and modalitity fusion. We believe that closer inspection of their embeddings and evaluation protocol is needed.

\section{Evaluation of two-tower systems}\label{sec:experiments_run}

\subsection{Dataset and models}\label{subsec:dataset_and_models}
We use the TinySOL dataset which contains 2913 audio clips with a single note played from a single instrument out of a set of 14 instrument classes. This dataset has been chosen as it has consistent recording settings without noise, it is a simple dataset for instrument recognition and finally, confounding factors (compression, sampling rate etc) are minimized.
We consider 3 models in total:
\begin{enumerate}
    \setlength\itemsep{0em}
    \item \textbf{Music/Speech CLAP}:~\cite{clap_learning_audio_concepts_elizalde_2023, laion_clap_Wu_2023} A CLAP-based model trained on music/speech data\footnote{music\_speech\_epoch\_15\_esc\_89.25.pt}
    \item \textbf{Music CLAP}:~\cite{laion_clap_Wu_2023} A CLAP-based model trained on music data\footnote{music\_audioset\_epoch\_15\_esc\_90.14.pt}
    \item \textbf{MusCALL}: A version of ~\cite{muscall_manco_2022}, retrained on music data\footnote{https://github.com/ilaria-manco/muscall}
\end{enumerate}

We use the two pretrained CLAP systems provided by LAION\footnote{https://github.com/LAION-AI/CLAP}.
For this work, the original MusCALL implementation was retrained from scratch, as both the data and trained models used in the original paper are not publicly available. 
Instead, we train on the LPMusicCaps-MTT~\cite{Doh2023LPMusicCapsLP} dataset, which is built by leveraging the audio and 188 tags from Magna Tag A Tune\cite{Law2009EvaluationOA} to artificially generate captions through a GPT-3.5 model.
The audio is resampled to 44.1 and 16 KHz for CLAP and MusCALL respectively, and the pre-processing steps described in their respective code repositories are followed.

\subsection{Zero-shot transfer for instrument classification}\label{subsec:problem_formulation}
Given an unseen audio segment $x^{*}$, a text label $l^{*}$ and a two-tower system $f(x)$, we want to model the likelihood $P(l^{*} | x^{*})$ based on the embeddings provided by $f(x)$. In the general case, $f \mapsto \mathbb{R}^{F}$ is a function that represents a two-tower system and maps audio or text information to a joint audio-text space, where $F$ is the dimension of the joint space. Also, let $\delta: (\mathbb{R}^{F} \text{ x } \mathbb{R}^{F}) \rightarrow \mathbb{R}$ be a function that measures similarity between joint space embeddings. In this approach, we model the $P(l^{*} | x^{*})$ based on $\delta$, as in:

\begin{equation}\label{eq:likelihood_approximation}
P(l^{*} | x^{*}) \propto \delta(f(x^{*}), f(l^{*}))
\end{equation}

Multiclass classification attributes the most probable class to each recording and equivalently, the one that has the maximum likelihood. Given our approximation, the output class for each recording is:

\begin{equation}\label{eq:multiclass_classification}
c^{*} = \argminB_{c \in C} \delta(f(x^{*}), f(c))
\end{equation}

for $c \in C$, where $C = \{c_{1}, c_{2}, \cdots, c_{N}\}$ is the set of classes that we are interested in.

In our work, the embedding similarity function $\delta$ is the cosine similarity:
\begin{equation}\label{eq:cosine_similarity}
\delta(e_{1}, e_{2}) = \frac{e_{1} 
\cdot e_{2}}{ \lVert e_{1}\rVert \cdot  \lVert e_{2} \rVert}
\end{equation}

where $\lVert\cdot\rVert$ is the $L_2$ norm and $e_{i} \in \mathbb{R}^F$ are embeddings.

\begin{figure*}[h!]
 \centerline{\includegraphics[width=0.9\linewidth]{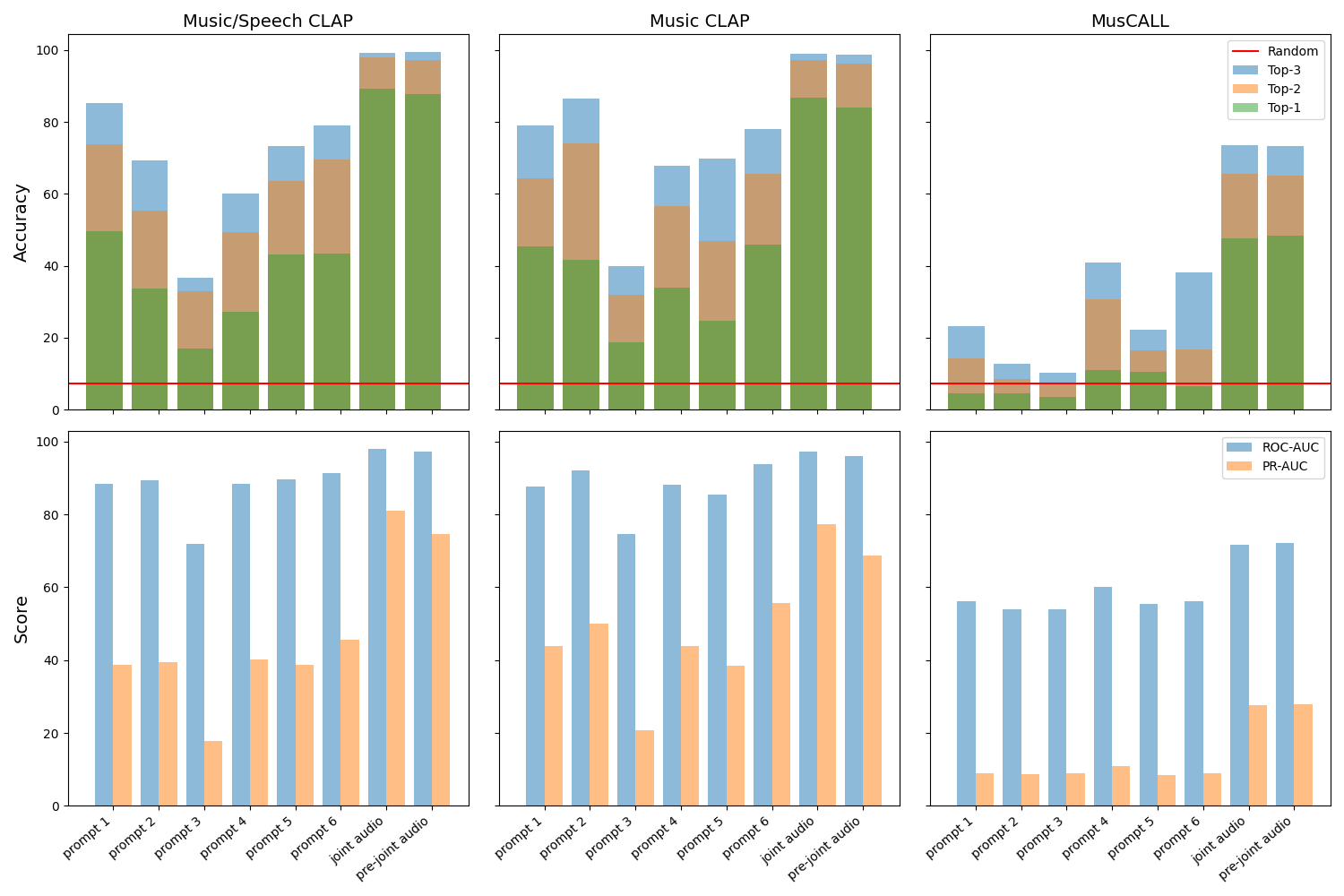}}
 \caption{Metrics for 6 textual prompts (See Section~\ref{subsec:are_two_tower_systems_context_dependent}), 2 audio based label embeddings (See Section~\ref{subsec:closely_inspecting_the_cosine_similarity_distribution}) and the 3 two-tower multimodal systems. The top row contains top-1 through top-3 accuracy and the bottom ROC-AUC and PR-AUC. The red line represents random choice.}
 \label{fig:prompt_evaluation}
\end{figure*}

\subsection{Experiment 1: Are two-tower systems context dependent?}\label{subsec:are_two_tower_systems_context_dependent}
Two-tower systems are typically not trained on single words, but rather longer prompts. As a result, the embedding produced for a single-word text label (e.g. ``guitar'') can be very different from the embedding for a longer prompt, with additional context (e.g. ``a guitar track''). When using two-tower models for classification, the class label can be wrapped in a prompt such as ``A <label> track''~\cite{muscall_manco_2022}, to better match the training distribution.  Methods to introduce stochasticity in the prompts used during training have been empirically proven to lead to more robust results~\cite{toward_universal_text_to_music_retrieval_doh_2023}. Retraining the systems and testing different ways of augmenting captions used for training is left for future work, but works in image-text~\cite{8953669} and video-text~\cite{modalit_balanced_embedding_wang_2022} two-tower systems provide some evidence for their usefulness.

The impact of different approaches for giving additional context to a single-word text label during inference has not been well-explored.
We explore the prompt sensitivity of each system by slightly changing the text prompt used for zero-shot classification in order to better understand to which extent these systems leverage contextual information. As far as we are aware, we are the first to evaluate the use of different types of prompts for two-tower systems during inference. Specifically, we evaluate 3 systems against 6 different prompts:

\begin{enumerate}
    \setlength\itemsep{0em}
    \item MusCALL prompt: ``\textit{A <label> track}''
    \item Generated definition: ``\textit{The <label> is a ...}''
    \item Generated definitions without label words: ``\textit{The <removed> is a ...}''
    \item Label word with random context: ``\textit{<label> <randomly selected lorem ipsum segment>}''
    \item Musically informed \#1: ``\textit{This is a recording of a <label>}''
    \item Musically informed \#2: ``\textit{Solo musical instrument sound of a <label>}''
\end{enumerate}

The first prompt proposed is the prompt that was used in MusCALL. The second prompt is generated using GPT-3.5~\cite{Brown2020LanguageMA}. The third prompt is the same as the second but we removed all instances of the label itself to evaluate the influence of the context on its own. To evaluate if the systems are sensitive to specific words and to further evaluate if the context is useful, the fourth prompt adds random words alongside the label. Lastly, we test 2 musically informed prompts.

As a first metric, we consider Top-k accuracy with $k = \{1,2,3\}$. We calculate the cosine similarity between each recording and instrument prompt and sort them. We assign zero-shot class labels as described in Section~\ref{subsec:problem_formulation} and check if the true label is present in the top-k assigned class labels. Furthermore, we calculate Receiver Operating Characteristic and Precision-Recall Area Under the Curve (ROC-AUC and PR-AUC respectively) following~\cite{won2020eval}.

Figure \ref{fig:prompt_evaluation} presents the zero-shot instrument recognition results for the three models across the 6 prompts, as well as the audio-only alternatives that will be described in Section \ref{subsec:closely_inspecting_the_cosine_similarity_distribution}. Despite the focus on music data, Music CLAP doesn't display very different results from Music/Speech CLAP. While music-specific systems are generally expected to perform better, this is not the case for two-tower systems. This might be an indication that music requires special treatment, as the metrics approach the state of the art in audio-text~\cite{laion_clap_Wu_2023} and image-text~\cite{NEURIPS2023_123a18df} two-tower systems.

Top-1 accuracy is worse than random for 4 out of 6 textual prompts for MusCALL. This might be caused by the small size of training data used, the absence of instrument-specific captions or their underrepresentation in the captions used, as well as the absence of single-note recordings in LPMusicCaps-MTT. While the metrics are low for MusCALL in most of the cases, a relatively large performance is still evident for the audio-only scenario. This implies that the problem lies in the audio-text alignment or the text branch.

The performance of CLAP models seems to be heavily correlated with the instrument labels themselves. Removing the label from definitions provides evidence that relevant context cannot be leveraged properly. Also, using musically informed prompts doesn't always result in greater or even comparable results. Specifically, top-1 accuracy drops when using the second musically informed prompt for CLAP models, despite the prompt being a more precise description of what is occurring in the audio.

These results suggest that CLAP models do not leverage extra context in the input prompt effectively. Both models performed worse when using relevant context without the instrument word, suggesting that the textual encoders put a lot more emphasis on the presence of specific words rather than the meaning of the prompts themselves. In addition, using a generic prompt provided better results than a musically informed one in most cases. Furthermore, any kind of context added at the prompts seems to harm the performance in most of the cases and provide more evidence that the model's text encoder cannot properly decompose the sentence to its constituents and use these semantically. Despite this observation, using definitions (prompt 2) seems promising for Music CLAP and for every metric apart from top-1 accuracy.

In the following experiments, we will consider only the ``MusCALL prompt'' as it leads to the highest top-1 accuracy, when model accuracy surpasses random choice.

\subsection{Experiment 2: Inspecting the cosine similarity distributions}\label{subsec:closely_inspecting_the_cosine_similarity_distribution}

\begin{figure}[h!]
    \centering
    \begin{subfigure}[b]{\columnwidth}
        \centering
        \includegraphics[width=\columnwidth]{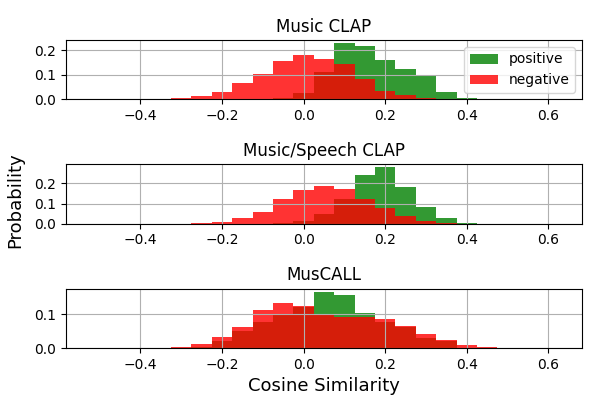}
        \caption{Histogram of cosine similarity between TinySOL data and MusCALL prompts in joint audio-text space.}
        \label{fig:positive_negative_histogram_text}
    \end{subfigure}
    \\
    \begin{subfigure}[b]{\columnwidth}
        \centering
        \includegraphics[width=\columnwidth]{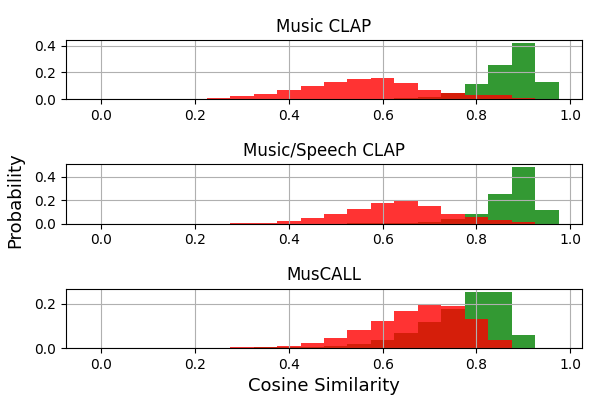}
        \caption{Histogram of cosine similarity between TinySOL audio data and the mean of intra-class embeddings in joint audio-text space.}
        \label{fig:positive_negative_histogram_audio}
    \end{subfigure}
    \caption{Histograms of audio and label embeddings for positive and negative pairs. When using textual prompts (\subref{fig:positive_negative_histogram_text}), the alignment is problematic, as can be seen from the overlap between positive and negative distributions.}
    \label{fig:similarity_histograms}
\end{figure}

As a next experiment, we calculate the cosine similarity between the joint space embeddings of each recording from TinySOL and the MusCALL prompt for each instrument, then compare the similarities of positive pairs vs negative pairs. We define a positive pair as an audio-label pair where the label corresponds to the instrument in the recording, and the negative pairs as all other pairs. Figure~\ref{fig:positive_negative_histogram_text} presents histograms of similarities for positive and negative pairs when using text prompts. If the audio-text coherence is good, positive and negative histograms should be well separated.

Positive and negative similarity distributions overlap greatly, as can be seen in Figure~\ref{fig:positive_negative_histogram_text}. As a result, retrieval is far from optimal. Fundamentally, a caption is a multi-faceted sentence. We suspect that treating a sentence as only one embedding point (mean of word embeddings) is fundamentally problematic and greatly hinders the semantic properties of the joint space. A hypothesis that needs testing is that by using composite sentences, a model cannot properly infer the relative embeddings of the sentence constituents.

\begin{figure}
 \centerline{\includegraphics[width=\columnwidth]{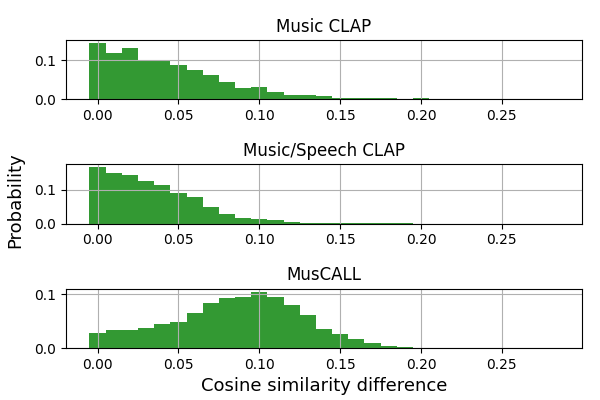}}
 \caption{The histogram of top-2 class similarities for every song in TinySOL. The CLAP models tend to be not very confident while the metrics are greater than the overconfident MusCALL with the worst metrics.}
 \label{fig:top_2_confidence_difference}
\end{figure}

To further evaluate if the audio encoder produces meaningful representations, we use the mean of intra-class song embeddings as the label embedding. This label embedding takes the role of the prompt embedding in the previous experiment. We generate the embeddings in joint audio-text space for each song. Then, we collect the songs that belong to k-th class $c_{k}$ and estimate the mean of the embeddings. The latter serves as the optimal embedding that the text label would have to be mapped to in order to maximize performance and will be referred to as the ``audio-only'' label. Note that this embedding is only optimal in the case of TinySOL data.

The resulting histograms for positive and negative pairs are shown in Figure~\ref{fig:positive_negative_histogram_audio}. They are well separated, indicating that the audio-encoder itself produces meaningful, separable embeddings.

Audio embeddings seem to be of good quality before and after the projection to the joint audio-text space, as the metrics are almost equal before and after projection in Figure~\ref{fig:prompt_evaluation}. The metrics almost double when using any audio-only labels, which further provides evidence that the problem resides in the text branch, or joint-space projection and there remains a large performance gap to bridge.

\subsection{Experiment 3: How confident are two-tower systems in their prediction?}

We calculate the histogram of the difference between the top-2 candidate classes for each recording~\cite{active_learning_survey_aggrawal_2014} to quantify the classification confidence. The similarity between each audio and instrument embeddings is estimated and they are sorted in descending order. The difference between top-2 similarities for each song is then calculated and a histogram of that difference is plotted in Figure \ref{fig:top_2_confidence_difference}. 

MusCALL seems to be overly confident in its prediction, which is unwarranted given the metrics reported. The opposite can be stated for CLAP models, where despite their better performance, the difference has a median value of 0.05-0.08.

\subsection{Experiment 4: Quantitative evaluation of the text branch}\label{subsec:the_problem_of_evaluating_the_text_branch}
While there are datasets that can be used to quantify the semantic properties and/or quality of a LM, there isn't one that focuses on music. To overcome this lack of text data for the case of instrument recognition, we can utilize instrument ontologies, which encompass semantic similarity of instruments at multiple levels. We propose to leverage them to quantify the semantic similarity between different instruments and instrument families. In this experiment we use the instrument ontology by Henry Doktorski\footnote{https://free-reed.net/} (HDIO). As every instrument ontology has its limitations~\cite{knowledge_representation_issues_instrument_kolozali_2011}, repeating the same experiment with other ontologies is left for future work.

We extract the tree based on HDIO and form every possible triplet of TinySOL instrument labels in the tree for a total of $364 = {14\choose3}$ combinations of positive word pairs without repetition. The triplets are of the form \emph{(<anchor>, <positive>, <negative>)} where the \emph{<anchor>} label has to be more semantically similar according to HDIO to the \emph{<positive>} than the \emph{<negative>}, e.g (\emph{``violin''}, \emph{``violoncello''}, \emph{``trumpet''}). Subsequently, every \emph{(<anchor>, <positive>)} pair that is linked through the root node of HDIO is excluded. The number of remaining triplets is 273.
As a way to quantify semantic meaningfulness with respect to musical instruments, we calculate cosine similarity between the \emph{(<anchor>, <positive>)} and \emph{(<anchor>, <negative>)} pairs for each system. Triplets for which the similarity is higher for the first pair than the second are considered ``correct'', and triplets where this is not the case are considered ``incorrect''. We compute the accuracy score as the percentage of correct triples.

We repeat this procedure with every valid triplet from the full ontology, as opposed to just using the instrument labels appearing in TinySOL. This gives us $\approx 443k$ triplets.

\begin{figure}
 \centerline{\includegraphics[width=0.9\linewidth]{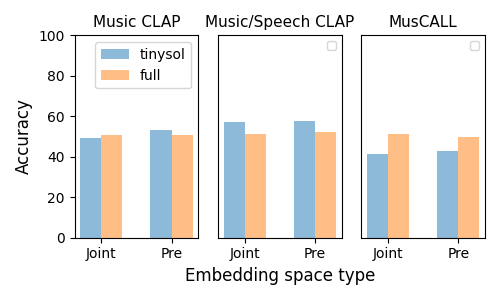}}
 \caption{Semantic meaningfulness quantification leveraging Henry Doktorski's instrument ontology. We evaluated the systems over valid triplets obtained through TinySOL labels, as well as every available triplet obtained from the ontology's labels. Accuracy ranges from 49-59\% which stresses that the models do not properly understand musical instruments in depth.}
 \label{fig:model_accuracy_by_subset}
\end{figure}

The accuracy for triplets from TinySOL and the full HDIO ontology are both presented in Figure~\ref{fig:model_accuracy_by_subset}. We see that half of the triplets are ``incorrect'' and this means that abstract semantic relations between instruments are not effectively captured in the textual branch, indicating a need for fine-tuning textual encoders on music related data.
Note that the accuracy is roughly the same as we would obtain by creating arbitrary triplets, though it is important to highlight that several instruments and instrument categories are words that are not frequently used in English.
Closer examination of the validity and usefulness of specific triplet cases (e.g. ``stringed'', ``plucked'', ``violin'') is left for future work.


\subsection{Experiment 5: Does joint space mapping introduce noise?}
To further examine the origins of the problematic embedding alignment, we repeat zero-shot evaluation with audio-only labels described in Sections~\ref{subsec:closely_inspecting_the_cosine_similarity_distribution} and semantic meaningfulness evaluation described in~\ref{subsec:the_problem_of_evaluating_the_text_branch} with the embeddings in the audio space and text space before the joint audio-text space mapping.

A minor performance increment can be seen when using the joint embedding instead of the pre-joint audio embedding, as can be seen in the last two columns of Figure~\ref{fig:prompt_evaluation}, apart from MusCALL where the metrics remain almost the same. We believe that the reduction in dimensionality of the joint space compared to the separate spaces is the underlying cause of these increments.

On the other hand, the accuracy based on HDIO remains the same, except for a decrement observed for Music CLAP and TinySOL subset of HDIO triplets, as can be seen in Figure~\ref{fig:model_accuracy_by_subset}. This could be an indication that the MLP can effectively map knowledge to the joint space. This is a further hint that potentially the problem lies in the LM used and fine-tuning might be needed to enforce musical semantics to be better represented.

\section{Conclusions and Future Work}
In this paper, we evaluated 3 two-tower multimodal systems for instrument classification. We provided a zero-shot classification analysis and an elaborate evaluation of the audio and text embeddings in the pre-joint and joint audio-text spaces. We also proposed a novel way to quantify the semantic meaningfulness of text embeddings based on triplets derived from an instrument ontology.

Generally, experiments showed that audio encoders are of good quality and hence, the alignment issue might be traced back to the text branch and/or the joint audio-text space mapping. Therefore, a solution could be to freeze the audio encoder and map the text information to audio space. Also, further attention to modality imbalance~\cite{Wang2022ModalityBalancedEF} can be placed with weighing in negative and positive examples~\cite{Onea2022ImprovingMS, Margatina2021ActiveLB, choiProperContrastiveSelfsupervised2022, muscall_manco_2022, DBLP:journals/corr/abs-2009-09805}. Additionally, to avoid sensitivity towards instrument labels and the inability to leverage context, we propose to use text augmentation over captions or masking/removing the words from them. It is important to state that the relation between sentence and word embeddings is not as straightforward as with bag-of-words Language Models~\cite{factors_affecting_sentence_similarity_Alian_2020} and as a result, the way to utilize captions or put additional emphasis on their constituents have to be further tested.

As a result, using two-tower systems might not be very useful for multi-class scenarios, given the large overlap between positive and negative histograms of cosine similarities shown in our experiments. We believe that it is essential for a music terminology similarity corpus to be established. The benefits will be two-fold: (1) it will provide a useful way of quantifying the semantic meaningfulness of the textual branch for two-tower model and (2) it can serve as a baseline to quantify the need for music-informed fine-tuning. Last but not least, genre and emotion ontologies can be used to further evaluate the semantic meaningfulness of language models.

\bibliography{ISMIRtemplate}

\end{document}